\documentclass{article}
\usepackage{graphicx} 
\usepackage{braket}
\usepackage{amsmath}
\usepackage{xcolor}
\usepackage{amssymb}
\usepackage[hidelinks]{hyperref}
\usepackage{authblk}
\usepackage[paperwidth=8.5in,paperheight=11in,centering,hmargin=2cm,vmargin=2.5cm]{geometry}
\usepackage{changes}
\usepackage{cite}
\newcommand{\RZ}{$R_Z(\phi)$ }
\newcommand{\RZnophi}{$R_Z$ }
\newcommand{\GR}{$R(\theta,\phi)$ }
\newcommand{\CZ}{CZ }
\newcommand{\CZnsp}{CZ}
\newcommand{\RZof}[1][$\phi$]{$R_Z(${#1}$)$}
\newcommand{\RXof}[1][$\theta$]{$R_X(${#1}$)$}

\title{A universal neutral-atom quantum computer with individual\\optical addressing and non-destructive readout}

\author[1]{A. G. Radnaev\footnote{{These authors contributed equally to preparation of the manuscript.}}}
\author[1]{W. C. Chung$^*$}
\author[1]{D. C. Cole$^*$}
\author[1]{D. Mason$^*$}
\author[4]{T. G. Ballance}
\author[2]{M. J. Bedalov}
\author[2]{D. A. Belknap}
\author[1]{M. R. Berman}
\author[2]{M. Blakely}
\author[1]{I. L. Bloomfield}
\author[1]{P. D. Buttler}
\author[3]{C. Campbell}
\author[4]{A. Chopinaud}
\author[1]{E. Copenhaver}
\author[1]{M. K. Dawes}
\author[1]{S. Y. Eubanks}
\author[1]{A. J. Friss}
\author[1]{D. M. Garcia}
\author[1]{J. Gilbert}
\author[2]{M. Gillette}
\author[3]{P. Goiporia}
\author[3]{P. Gokhale}
\author[2]{J. Goldwin}
\author[2]{D. Goodwin}
\author[5]{T. M. Graham}
\author[1]{CJ Guttormsson}
\author[1]{G. T. Hickman}
\author[2]{L. Hurtley}
\author[1]{M. Iliev}
\author[1]{E. B. Jones}
\author[1]{R. A. Jones}
\author[1]{K. W. Kuper}
\author[1]{T. B. Lewis}
\author[2]{M. T. Lichtman}
\author[1]{F. Majdeteimouri}
\author[2]{J. J. Mason}
\author[1]{J. K. McMaster}
\author[2]{J. A. Miles}
\author[1]{P. T. Mitchell}
\author[2]{J. D. Murphree}
\author[2]{N. A. Neff-Mallon}
\author[1]{T. Oh}
\author[3]{V. Omole}
\author[1]{C. Parlo Simon}
\author[2]{N. Pederson}
\author[3]{M. A. Perlin}
\author[1]{A. Reiter}
\author[3]{R. Rines}
\author[2]{P. Romlow}
\author[1]{A. M. Scott}
\author[1]{D. Stiefvater}
\author[1]{J. R. Tanner}
\author[1]{A. K. Tucker}
\author[1]{I. V. Vinogradov}
\author[1]{M. L. Warter}
\author[1]{M. Yeo}
\author[2, 5]{M. Saffman}
\author[1]{T. W. Noel\footnote{Corresponding author tom.noel@infleqtion.com}}

\affil[1]{Infleqtion, Boulder, CO, USA  }
\affil[2]{Infleqtion, Madison, WI, USA  }
\affil[3]{Infleqtion, Chicago, IL, USA  }
\affil[4]{Infleqtion UK, Oxford, UK}
\affil[5]{Department of Physics, University of Wisconsin-Madison, Madison, WI, USA.}

\date{August 2024}
\begin{document}

\maketitle

\begin{abstract}

Quantum computers must achieve large-scale, fault-tolerant operation to deliver on their promise of transformational processing power~\cite{Shor1994, McArdle2020, Gao2022, Santagati2024}. This will require thousands or millions of high-fidelity quantum gates and similar numbers of qubits~\cite{Hoefler2023}. Demonstrations using neutral-atom qubits trapped and manipulated by lasers have shown that this modality can provide high two-qubit gate (\CZnsp) fidelities and scalable operation~\cite{Evered2023, Bluvstein2024, peper2025, Finkelstein2024, Wu2024, muniz2024high, rodriguez2024experimental, reichardt2024logical}. However, the gates in these demonstrations are driven by lasers that do not resolve individual qubits, with universal computation enabled by physical mid-circuit shuttling of the qubits. This relatively slow operation may greatly extend runtimes for useful, large-scale computation. Here we demonstrate a universal neutral-atom quantum computer with gate rates limited by optical switching times, rather than shuttling, by individually addressing tightly focused laser beams at an array of single atoms. We achieve CZ fidelity of 99.35(4)\% and local single-qubit \RZnophi gate fidelity of 99.902(8)\%. Moreover, we demonstrate non-destructive readout of alkali-atom qubits with 0.9(3)\% loss, which boosts operational speed. This technique also enables us to measure a state-of-the-art CZ fidelity of 99.73(3)\% when excluding atom-loss events, which may be mitigated through erasure conversion. Our results represent a critical step towards large-scale, fault-tolerant neutral-atom quantum computers that can execute computations on practical timescales. 

\end{abstract}

\section{Introduction}
 Quantum computers promise to reduce computational energy consumption and solve classically expensive or intractable problems in a range of fields including material science, machine learning, security, chemistry, energy, finance, medicine, and pharmaceutical development~\cite{Shor1994, McArdle2020, Gao2022, Santagati2024, Date2021, Ma2020, Koretsky2021, Perlin2024, Ramesh2024}. Recent analyses suggest that realization of practical utility will require fault-tolerant operation~\cite{Hoefler2023}. This in turn requires quantum error correction (QEC), which comes with large overhead in the number of qubits and requires high one- and two-qubit quantum gate fidelities. The field's progress along this path has entered a phase of experimental demonstrations~\cite{GoogleAI2023, Silva2024, Bluvstein2024, rodriguez2024experimental, reichardt2024logical, bedalov2024fault}. Among the diverse modalities proposed for large-scale fault-tolerant quantum computation~\cite{Bergou2021fromSaffman}, the neutral-atom approach is one of the most promising candidates and has become a leader in the combination of qubit count, universal gate-based circuit operation, connectivity, and quantum error correction~\cite{Bluvstein2024, Manetsch2024, Graham2022, rodriguez2024experimental, reichardt2024logical}. The rapid progress of neutral atoms for quantum computing can be attributed to the combination of naturally identical qubits; long coherence times due to decoupling from the environment~\cite{Barnes2022}; switchable qubit interaction strengths across 12 orders of magnitude~\cite{Saffman2010}; sub-microsecond, high-fidelity, scalable gates~\cite{Evered2023, Jandura2022, muniz2024high}; and the ability to trap and manipulate large, dense, multi-dimensional qubit arrays~\cite{Norcia2024, Manetsch2024, Huft2022}, as was recently highlighted in demonstrations of error-corrected logical qubits~\cite{Bluvstein2024, rodriguez2024experimental, reichardt2024logical}. 

\begin{figure}
    \centering
    \includegraphics[width=1.0\textwidth]{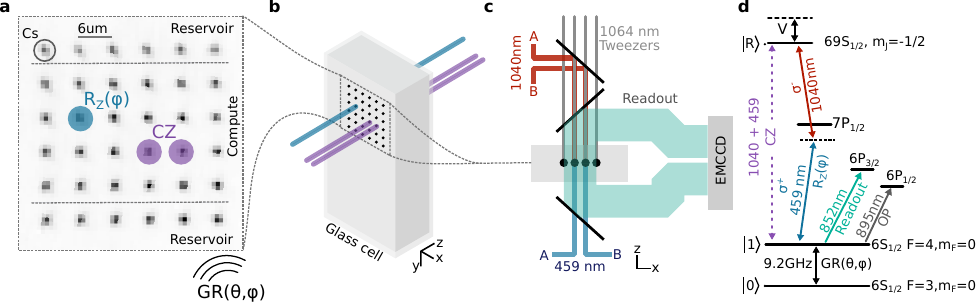}
    \caption{\textbf{System overview.} \textbf{a,} Image of array of Cs atoms, trapped in a glass cell. \textbf{b,} Locally addressed \RZ (beam shown in blue) and \CZ (beams show in purple) gates are applied via beams propagating perpendicular to the atom plane.  Microwave-based global \GR gates (labeled GR) address the entire array. \textbf{c,} Layout of key lasers implementing trapping and gates.  Rydberg excitation is implemented by counter-propagating 459~nm (blue) and 1040~nm (red) lasers.  Rydberg illumination for control and target qubits are sourced from separate pulse shaping systems (labeled A and B).  An optical tweezer array (1064~nm) is generated via AODs and combined with the 1040~nm light. Readout fluorescence is collected and imaged on an electron-multiplying CCD (EMCCD).  \textbf{d,} Atomic levels and addressing fields, including lasers for two-photon Rydberg excitation, readout, and optical pumping (OP). Rydberg beams are circularly polarized ($\sigma^+$ and $\sigma^-$ for 459~nm and 1040~nm, respectively).  Rydberg blockade shift denoted by $V$.}
    \label{fig:qpu_diagram}
\end{figure}

Two architectural approaches have emerged for neutral-atom quantum processors.  One \cite{Bluvstein2024, rodriguez2024experimental, reichardt2024logical} relies on simultaneously shuttling multiple qubits to and from a single, dedicated zone for implementing gates.  This allows flexible qubit connectivity and avoids the need for near-diffraction-limited addressing beams but increases compilation complexity~\cite{Tan2024} and limits gate rates (number of gates per unit of time within a circuit) due to constraints on the speed of atom movement\cite{Lam2021}. 
High gate rates are important for maintaining practical computation times, particularly in light of the physical qubit count and gate depth overheads required for fault-tolerant operation.
An alternative approach \cite{Graham2022} keeps the qubits stationary and addresses them individually with steerable laser beams, enabling gate rates limited only by optical switching, which is inherently faster than atom movement. This stationary qubit approach poses challenges for achieving high gate fidelities and does not natively support all-to-all connectivity, but it offers the promise of faster circuits, in direct support of long-term fault-tolerant operations \cite{Poole2024}.  Here, we present a full-stack neutral-atom quantum computer using gates based on individual optical addressing instead of atom motion.  Overcoming the challenges specific to this architecture, we achieve entangling gate fidelity of 99.35(4)\%, already exceeding QEC thresholds for certain codes~\cite{Knill2005, Sahay2023, Bravyi2024, Hong2024}. Our results open the door to faster approaches for fault-tolerant operation that exclusively use individual optical addressing of stationary qubits, or that combine this with some atom shuttling to optimize execution of error-corrected circuits.  We further advance the state-of-the-art by incorporating non-destructive qubit readout, enabling faster computations through qubit re-use.

\section{A neutral-atom architecture with individual optical addressing}\label{sec: quantum_computer_architecture}

Building on prior work \cite{Graham2022}, we present a full-stack neutral-atom quantum computer using an individual optical addressing architecture.  Arbitrary quantum circuits can be submitted via the cloud, where they are compiled\footnote{via Infleqtion Superstaq~\cite{Campbell2023}} into a universal native gate set consisting of global microwave \GR rotations, arbitrary local \RZ rotations, and nearest-neighbor controlled-Z (\CZnsp) gates. The qubits are optically trapped Cs atoms, with $\ket{0}$ and $\ket{1}$ encoded in the $\ket{F=3, m_F=0}$ and $\ket{F=4, m_F=0}$ clock states of the ground state hyperfine manifold, respectively.  The atoms are trapped out of a cold atomic beam\footnote{Infleqtion PICAS}, cooled, and rearranged by an optical tweezer to form a defect-free array.  Fig. 1 illustrates a 24-qubit computational array alongside 12 reservoir atoms, formed in a 1064 nm tweezer array (see Methods Sec. \ref{sec:qpu}). After state preparation via optical pumping, arbitrary circuits are implemented with microwave and laser pulses, followed by non-destructive state-selective readout (NDSSR) of the qubit state. During this readout the qubits remain trapped in the optical tweezers with high probability, allowing re-initialization and re-use of the same atomic array for the next shot (see Sec. \ref{sec:NDSSR}).  After the requested number of measurements is collected, circuit results are returned to the user. Throughout this process feed-forward stabilization is applied to background electric and magnetic fields, and lasers are stabilized on various timescales using both optical power measurements and atom-based signals.  

Unique to this system is the dual-beamline optical design supporting laser-based local \RZ rotations and \CZ gates.  The former requires application of 459~nm light to a single atom, and the latter requires simultaneous application of 459~nm and 1040~nm light to a pair of atoms to drive two-photon Rydberg excitation.  To implement these gates, we send each laser through a sequence of three acousto-optic devices:  one for pulse shaping and two for beam steering. The pulsed and steered beams are then focused to nearly diffraction-limited spots on the atom array (see Fig. \ref{fig:qpu_diagram}).  Using both direct optical measurements and atom-based beam profiling, we confirm that our 459 and 1040 beams are focused to spot sizes of 2.8 $\mu\rm m$ and 3.0 $\mu\rm m$ ($1/e^2$ intensity radii) in the atom plane. Critically, these spots are smaller than the atom spacing of 6 $\mu\rm m$, allowing us to address all qubits individually.  With this approach, we can alternate between \CZ and local \RZ gates (on select target qubits) with a timescale limited only by the beam-steering devices.  Our current acousto-optic deflectors (AODs) used as beam-steering devices can be re-aimed with a timescale of 2 $\mu\rm s$. In future devices, driving AODs with multi-tone RF signals and/or combining AODs with spatial light modulators (SLMs) can enable fast, individually addressed gates on multiple qubits simultaneously \cite{Graham2023}. With this approach our individual-addressing architecture supports parallel gate operations while maintaining high gate rates.

A practical consideration for this individual addressing design is the crosstalk that can occur from residual beam intensity reaching a neighboring atom.  For instance, given our beam waists, one predicts 0.03\% intensity reaching neighboring qubits. In practice, we find crosstalk intensities up to ${\sim}10$x higher, likely due to optical aberrations. We provide a thorough discussion of the implications of this crosstalk in Methods Sec. \ref{sec:crosstalk_simulations}, and summarize the results here.
One effect of this crosstalk is residual light reaching spectator atoms not participating in the gate.  For our current parameters, we estimate that this produces an error contribution smaller than $10^{-4}$ per gate. Moreover, it is a predictable unitary error that can be compiled away \cite{Campbell2023}.  The other important impact of crosstalk is on the entangling gate itself, where interference between the control and target illumination can significantly perturb the gate dynamics.  However, we find that this error contribution can be made negligible by ensuring that the fields interfere at a frequency much higher than the Rabi frequency of the Rydberg excitation.  
We design our addressing system to satisfy this requirement. For each wavelength, we branch the single source laser into two fully independent beam-shaping and beam-steering paths (labeled A and B in Fig. 1). Thus, in a \CZ gate, the Rydberg laser light reaching the control and target qubits have disparate acousto-optic frequency shifts, resulting in controllable crosstalk beatnote frequencies in the range $0 - 30$ MHz (see Methods Sec.  \ref{sec:crosstalk_simulations}). This dual-beamline system enables our processor to tolerate greater crosstalk intensity at higher array density, achieving higher blockade strength, while still addressing qubits individually and minimizing the impact of crosstalk on neighboring atoms.

\section{Quantum processor implementation and characterization\label{sec:implementation}}
Optically trapped Cesium atoms offer a ground-state hyperfine clock qubit with excellent  coherence properties. We find a trap vacuum lifetime of 26(2) s, limited by background gas collisions. Raman sideband cooling is used to produce an atom temperature of 2.6(3) $\mu$K, as indicated by both release-recapture measurements\cite{Tuchendler2008} and a Doppler-limited $T_2^*=15(1)$ $\mu s$ for the ground-Rydberg qubit. This temperature alone would permit a clock-qubit coherence time of about 42 ms\cite{Kuhr2005}. In practice we measure $T_2^*=12.7(5)$ ms, and we attribute this additional decoherence to magnetic field shot-to-shot fluctuation of ${\sim}1$~mG at our bias of 11.15~G, as supported by measurements of $T_2^*<1\,\text{ms}$ on magnetically sensitive hyperfine transitions. By concatenating repeated XY8 dynamical decoupling sequences, we can extend $T_2$ beyond the capability of our hardware to accurately measure it---our best estimate is 2.8 s, with a 97\% confidence interval lower bound of 128 ms.

\begin{figure}
    \centering
    \includegraphics[width=89mm]{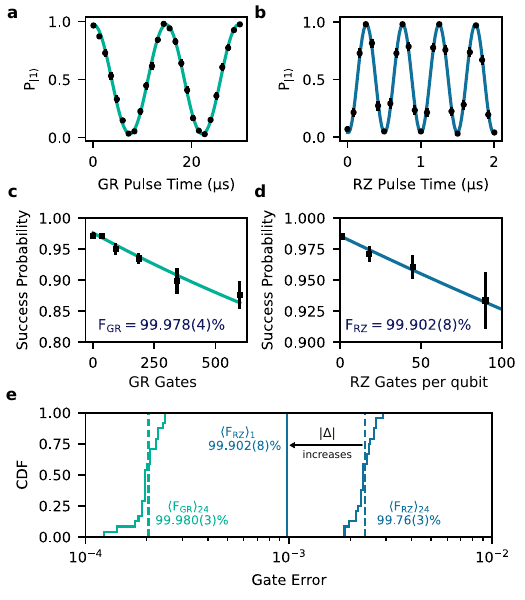}
    \caption{\textbf{Single-qubit gates.} \textbf{a,} Rabi oscillations driven by a global microwave field. \textbf{b,} Ramsey oscillations from local qubit \RZ control. Error bars in (a, b) represent projection noise. \textbf{c, d,} Randomized benchmarking results for global \GR and local \RZ gates, respectively. Each point is an average over multiple random circuits (10 per depth for \GR and 5 per depth for \RZ), with error bar given by standard error of the mean. \textbf{e,} Randomized benchmarking results, plotted as cumulative distribution function (CDF) of 24-qubit (or 1-qubit) results. For 24-qubit results, gates are applied to each of the 24 qubits in each circuit. For the single qubit results, the reported uncertainty is the benchmarking fit uncertainty.  For the 24-qubit results, the reported uncertainty is the ensemble standard deviation.}
    \label{fig:gr_rz_benchmarking}
\end{figure}
Our single-qubit gate set consists of global \GR (GR) and local \RZ rotations.  
The former are implemented via a uniform pulsed microwave field resonant with the magnetically insensitive qubit $\ket{0}\rightarrow\ket{1}$ transition.  
The duration and phase of the pulse determines $\theta$ and $\phi$, respectively.  
Fig. \ref{fig:gr_rz_benchmarking}a illustrates Rabi oscillations driven by this microwave field.  
With this system, we have reached Rabi rates as high as $\Omega_\mathrm{GR}$ $=2\pi \times 120$~kHz, enabling an $R(\pi, \phi)$ gate time of 4.1~$\mu$s.
We characterize the full set of GR Cliffords across a 24-qubit system using randomized benchmarking (Fig. \ref{fig:gr_rz_benchmarking}c, e), and find an average fidelity of 99.980\%, with an ensemble standard deviation of 0.003\% and a typical fit uncertainty of 0.004\%.  

Local \RZ rotations are implemented via the differential light shift arising from a locally addressed, 459~nm laser beam that off-resonantly addresses the transition from $\ket{1}$ to the $7P_{1/2}$ excited state with a detuning  $\Delta$ from the $7P_{1/2}$ center of mass.
Fig. \ref{fig:gr_rz_benchmarking}b presents Ramsey measurements of a differential light shift of $2 \pi \times 2$~MHz, enabling an \RZof[$\pi$] gate time of 250 ns. We characterize \RZ fidelity for two values of $\Delta$ using interleaved randomized benchmarking (IRB), with rotation angles sampled uniformly over $[0, 2\pi)$ (see Methods Sec. \ref{sec:rz_characterization}).
We first characterize the full 24-qubit array with $\Delta/2\pi=885$ MHz (and, incidentally, a reduced \RZ rate of $2 \pi \times 720$~kHz). This characterization uses a maximum IRB depth of $200$ \RZ gates per qubit, which entails redirecting the beam-steering AODs $4800$ times in a 15~ms circuit. We find an average \RZ fidelity of 99.76\%, with an ensemble standard deviation of 0.03\% (Fig. \ref{fig:gr_rz_benchmarking}e). To properly account for the difference between leakage and true depolarization, the fidelity calculation includes an estimate of the scattering error due to non-zero population of the $7P_{1/2}$ intermediate state (Methods Sec. \ref{sec:rz_characterization}).  Our observed \RZ error is dominated by this scattering, which can be suppressed by using larger detuning.  Indeed, changing $\Delta/2\pi$ to $-2.1$~GHz (and operating with an \RZ rate of $2 \pi \times 2$~MHz), we measure, for a single qubit, a higher \RZ fidelity of 99.902(8)\% (see Fig. \ref{fig:gr_rz_benchmarking}d). The residual error of 0.098(1)\% is largely explained by the remaining leakage probability of 0.066\% and laser pulse energy fluctuation.

\begin{figure}[t!]
    \centering
    \includegraphics[width=89mm]{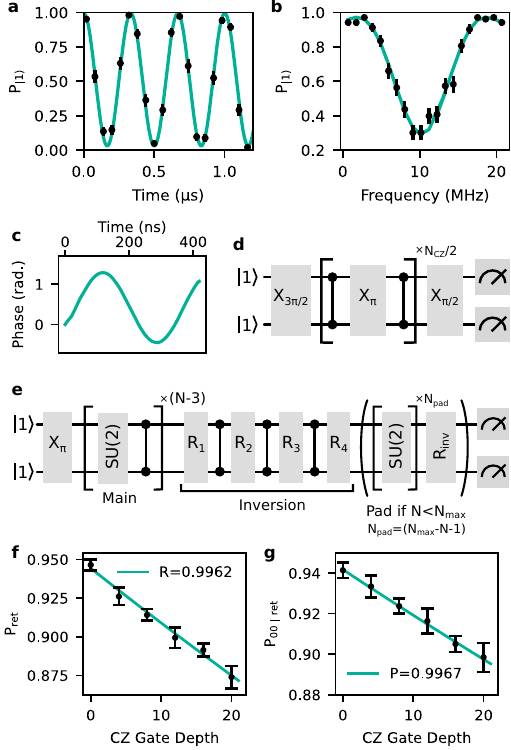}
    \caption{\textbf{Characterization of \CZ gates.} \textbf{a,} Rabi oscillations between $\ket{1}$ and $\ket{R}$, at a rate of $2 \pi \times 3$~MHz.  \textbf{b,}~Rydberg spectroscopy, measured while a neighboring atom is already in the Rydberg state.  Decreased qubit $\ket{1}$ population indicates simultaneous excitation to the Rydberg state, which is only possible when the excitation is detuned (x-axis) by the Rydberg blockade. Error bars in (a, b) represent projection noise. \textbf{c,} Phase profile of 1040~nm optical pulse used in the \CZ gate, measured using optical heterodyne interferometry.  \textbf{d,} Circuit used as cost function for \CZ gate optimization. Global $R(\theta,\phi=0)$ gates denoted as $X_\theta$. \textbf{e,} Circuits used for characterization of \CZ gates include a main block of $(N-3)$ pairs of alternating Haar-random global SU(2) gates and \CZ gates, as well as an inversion operation (consisting of three \CZ gates and four global rotation gates $R_i$) and a padding circuit that reduces the spurious impact of \GR errors on extracted CZ fidelity. For circuits with $N=0$ the ``Main" and ``Inversion" blocks are omitted entirely. \textbf{f,} Observed probability to retain both atoms as a function of the number of \CZ gates ($N$) in the circuit, along with the fit used to extract the retention probability. Decay is fit to an asymptote of zero. \textbf{g,} Observed probability to obtain the expected state-selective measurement outcome, given that both atoms were retained, as a function of the number of \CZ gates in the circuit, along with the extracted cycle polarization. Decay is fit to an asymptote of 25\%. Data in \textbf{f, g} are obtained from sequential NDSSR (\textbf{g}) and occupancy readout (\textbf{f}) measurements on the same quantum state. Each point is obtained as an average over between 18 and 48 circuits (183 total circuits in dataset), with an error bar corresponding to standard error of the mean.}
    \label{fig:cz_benchmarking}
\end{figure}

Our quantum processor implements the two-qubit controlled-Z (\CZnsp) gate, which, combined with arbitrary single-qubit rotations, is sufficient for universal quantum computation. 
 The gate is based on simultaneous excitation of the control and target qubits to a Rydberg state, where blockade interactions provide inter-qubit coupling.  
 In particular, we use two-photon excitation (459~nm + 1040~nm) to couple the $\ket{1}$ state to the $\ket{69S_{1/2}, m_j=-1/2}$ Rydberg level. The single-photon intensities are defined by the $\Delta/2\pi=-2.1$~GHz detuning from the $7P_{1/2}$ intermediate state, the $2\pi\times2$ MHz light shift applied to the qubit by the 459 nm laser and the $\Omega =2 \pi \times 3$~MHz Rabi frequency of Rydberg excitation (Fig. \ref{fig:cz_benchmarking}a). We measure a blockade interaction energy of $2 \pi \hbar \times 10.3(1)$~MHz (Fig. \ref{fig:cz_benchmarking}b).  We implement an approximation of the time-optimal Rydberg \CZ gate \cite{Jandura2022} using a pulse of duration 7.85/$\Omega$ = 416 ns with a sinusoidal phase profile (implemented via electro-optic modulation of the 1040~nm beam; see measured phase profile in Fig. \ref{fig:cz_benchmarking}c). The full \CZ gate consists of this phase-modulated Rydberg pulse followed by single-qubit phase corrections to produce the canonical gate.  We first calibrate the uncorrected gate, using a cost function that is insensitive to the single-qubit phases \cite{Evered2023} (Fig. \ref{fig:cz_benchmarking}d), then calibrate the phase corrections (see Methods Sec. \ref{sec:cz_calibration}). Note that this gate modulation is implemented upstream of the beam-steering AODs, so our individual-addressing architecture does not impose any extra calibration overhead for this protocol.

We characterize our \CZ gates using randomized benchmarking circuits in which \CZ gates alternate with global single-qubit gates (see Fig. \ref{fig:cz_benchmarking}e). 
Symmetric benchmarking is discussed in detail in Ref. \cite{Baldwin2020}, and approaches similar to ours were used in Refs. \cite{Evered2023, Ma2023, peper2025}.
This approach has limitations; for example, swap errors go completely undetected, and it strictly only characterizes the performance of the gate on a subset of two-qubit input states. Nevertheless, it is known to perform well given the error profile of neutral-atom platforms. To validate application of the approach to our gate, we perform density matrix simulations of our benchmarking procedure using a realistic microscopic error model for the gate (see Methods Sec. \ref{sec:benchmarking_simulations}) and find agreement between the fidelity extracted from benchmarking and the true underlying gate fidelity, reproducing findings from prior applications of the technique \cite{Evered2023, Ma2023}. 

Figure \ref{fig:cz_benchmarking} presents \CZ benchmarking results for a single pair of atoms.  In these circuits single-qubit phase corrections are implemented via a combination of local \RZ and virtual phase shifts to the local oscillator, as in Ref. \cite{Tsai2024}. 
Because we characterize a single qubit pair, the beam-steering AODs are statically driven throughout the control sequence.
By following each NDSSR measurement immediately with an occupancy measurement, we are able to separately characterize the two-atom survival probability ($R=99.62(3)$\% per gate) and cycle polarization ($P=99.67(4)$\%).  
Combining these with a model-based estimate of the leakage probability (see Methods Sec. \ref{sec:cz_benchmarking_details}), we extract a \CZ fidelity of $99.35(4)$\%.
The leading physical mechanisms contributing to the observed $0.65$\% \CZ gate error are leakage due to the finite lifetime of the $7P_{1/2}$ intermediate state, leakage and loss due to the finite lifetime of the Rydberg level, and shot-to-shot intensity fluctuations of the Rydberg laser beams.  We estimate the contribution from optical crosstalk to \CZ error to be $\lesssim 0.1$\% in our current configuration, with a straightforward path to further reduction (see Methods Sec. \ref{sec:crosstalk_simulations}).

NDSSR allows us to filter the data based on atom retention and calculate a post-selected \CZ fidelity of 99.73(3)\% (see Methods Sec. \ref{sec:cz_benchmarking_details}). This ${\sim}0.38$\% atom loss error arises from population that remains in a Rydberg state at the end of the gate, as indicated by separate measurements that show that loss from the temporary extinction of the tweezers during the gate is negligible. Atom loss error may be converted to erasure through a leakage-detection measurement in which an ancilla becomes entangled with the presence/absence of an atom in a qubit state (rather than its logical state)~\cite{Preskill1998, Chow2024}, enabling the use of erasure-specific techniques to achieve improved correction performance~\cite{Wu2022}.

\section{Non-destructive State Selective Readout\label{sec:NDSSR}}
While neutral-atom platforms have demonstrated significant improvements in gate fidelity and qubit count scaling that make neutral atoms a leading modality for quantum computing, there has been relatively less emphasis on their shot rate. Non-circuit operations such as atom loading and arrangement are expected to be a non-negligible part of the total operation time for measurement schemes based on destructive readout; a non-destructive readout scheme with repetitive measurements can significantly increase shot rate by amortizing the cost of these non-circuit operations and can enable mid-circuit measurements for fault-tolerant operation. 

\begin{figure}[p]
    \centering
    \includegraphics[width=89mm]{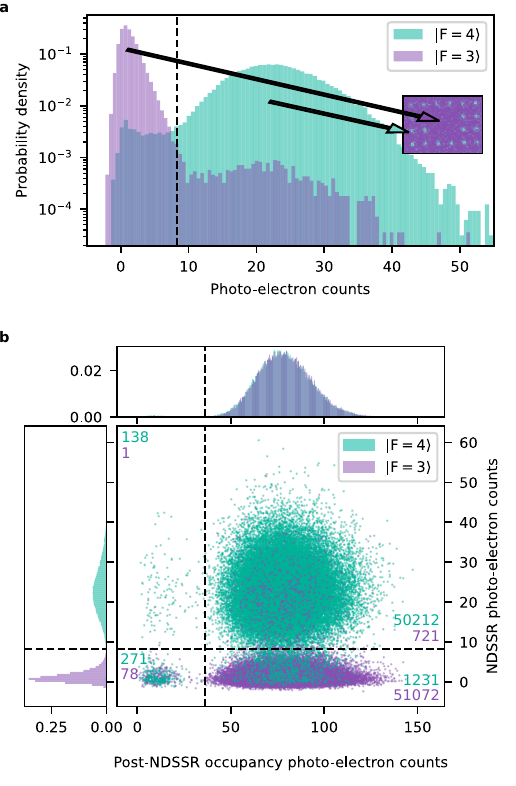}
    \caption{\textbf{Non-destructive state-selective readout.} NDSSR signal distribution from 43 repetitions of SPAM characterization experiment. Each experiment executes two circuits - empty circuit and circuit with single $R(\pi,0)$ gate - with approximately 600 shots each, resulting in approximately 25800 camera frames for each circuit. The turquoise-colored distribution describes the signals where atoms are expected to be in a bright state ($|F=4\rangle$) and the purple-colored distribution describes the signals for a dark state ($|F=3\rangle$). The samples for each distribution come from the deterministically loaded diagnostic site pair used for \CZ gate fidelity benchmarking. Photo-electron counts for each site were calculated from the raw EMCCD camera analog-digital units using nominal EM gain and pre-amplifier gain. Each quadrant of the plane defined by NDSSR and post-NDSSR photo-electron counts contains two text labels showing total counts from $|F=4\rangle$ preparation and measurement experiment (turquoise) and counts from $|F=3\rangle$ preparation and measurement experiment (purple). \textbf{a,} Sampled probability distribution of photo-electron counts for bright and dark states. The dashed black line denotes the discrimination threshold that yields discrimination fidelity of 99.6(2)\%. The shaded areas under the distributions on both sides of the discrimination threshold denote signals associated with state detection error. Depumping and atom loss during the NDSSR process and microwave pulse infidelity during state transfer from the stretched state to the bright qubit state give rise to bright state detection error of 3.1(7)\%, which is reduced to 2.6(7)\% when atom loss is post-selected out. The dark state detection error of 1.6(5)\% is believed to be a combination of (partial) state preparation error and $F=3 \rightarrow F=4$ repumping during imaging from residual leakage of repump laser intensity. The middle inset shows a single NDSSR measurement of a four-qubit GHZ state realized in the 24-qubit array.
    \textbf{b,} NDSSR and post-NDSSR occupancy readout photo-electron counts. The bright state atom loss rate is 1.0(4)\%  and the dark state atom loss rate is 0.8(3)\%. The atom loss rates and state detection errors and their uncertainties provided are the means and standard deviation of their measured values for the 43~SPAM characterization experiments that comprise the dataset.
    }
    \label{fig:ndssr_main_figure}
\end{figure}
We implement the non-destructive state detection method demonstrated in Ref.~\cite{Kwon2017} with a few critical improvements (see Methods~\ref{sec: ndssr_methods}). During the state detection sequence, qubit state $|1\rangle$ is optically pumped to the stretched state $\ket{F=4, m_F=4}$, where we can detect fluorescence from the closed $\ket{F=4, m_F=4} \rightarrow  \ket{F'=5, m_F'=5}$ transition, while $\ket{0}$ remains dark. By strategically alternating between illumination with trap light and readout light, we mitigate the deleterious effects of excited state light shift and dipole force fluctuations, and we reach a state-averaged loss rate of $0.9(3)\%$ (1.0(4)\% for the bright state and 0.8(3)\% for the dark state) while achieving bright-dark discrimination fidelity of 99.6(2)\% with 6 ms of camera exposure time. The imaging performance and the state-discrimination fidelity of the readout scheme is analyzed from a large dataset of images of the diagnostic site pair used for \CZ benchmarking analysis collected during state preparation and measurement (SPAM) characterization experiments, each of which prepares all qubits to be in $|1\rangle$ (bright) or $|0\rangle$ (dark) (Fig.~\ref{fig:ndssr_main_figure}). The raw state-discrimination fidelity without state preparation or atom loss correction is 97.7(5)\%, limited by bright state detection error of 3.1(7)\% and dark state detection error of 1.6(5)\%. Note that state preparation error, which includes failure to optically pump into $|F=4, m_F=4\rangle$ during Raman sideband cooling and failure to coherently transfer $|F=4, m_F=4\rangle$ to $|F=4, m_F=0\rangle$ using microwave pulses (see Sec.~\ref{sec:cooling_and_state_prep}), is distributed across both types of state detection errors. The shot rate enhancement from repetitive non-destructive measurements depends on the quantum circuit being run, especially on the number of \CZ gates as they can induce small atom loss. For an example circuit, we chose the preparation of a 15-qubit Greenberger–Horne–Zeilinger (GHZ) within a 24-qubit array, which contains 14 \CZ gates. We recorded an average of 8.2(2)~Hz shot rate for the example circuit compared to an estimated single-measurement shot rate of ${\sim}3$~Hz (see Sec.~\ref{subsubsec: shot_rate_enhancement}). These experiments were performed before the integration of Raman sideband cooling, which adds additional time to each measurement that cannot be amortized (see Methods Fig.~\ref{fig:ndssr_control_flow}).

The non-destructive readout method for alkali atomic species demonstrated in this work provides the starting point for repetitive mid-circuit readout of ancilla qubits in dual-alkali-species-based logical qubit architectures. The atom loss rate can be further reduced by using shorter exposure times and advanced camera image binarization algorithms that work at low photon counts~\cite{Su2024}. The depumping error during imaging can be improved with technical improvements in polarization purity and with microwave-assisted state-selective repumping during imaging~\cite{GrahamMidcircuit}.

\section{Outlook}

This work advances the state-of-the-art for individually addressed entangling gates in a neutral-atom platform, and we foresee a clear path towards the error levels required for fault tolerance.  Control errors may be reduced by reaching lower atom temperatures, better stabilizing laser power and frequency, and using robust gate protocols\cite{Jandura2023, Mohan2023}. Straightforward changes such as increasing the intermediate-state detuning will immediately reduce scattering error. Moving forward, scattering error can be eliminated by transitioning to single-photon Rydberg excitation \cite{Tsai2024, Ma2023}, which also reduces the complexity of light shifts on atomic resonances. For larger systems, parallel gate operations using multiple beams is possible through multi-tone AOD drives and/or SLMs. The path for scaling our qubit array is also well-established, with demonstrations\cite{manetsch2024tweezerarray6100highly} suggesting the feasibility of trapping $>10^4$ qubits in a single array, with the option to improve trapping lifetimes by operating at cryogenic temperatures\cite{PhysRevA.106.022611}.  The non-destructive readout we present here already enables state-of-the-art shot rate, and paths towards even faster readout are being developed\cite{PhysRevA.102.053101, Chow2023}. Finally, we note that our ability to achieve high ground-Rydberg coherence ($T_2^*=15~\mu\text{s}$) at $n=69$ is a unique advantage in the field, enabled by our electric field control.  Since the Rydberg interaction scales with $n^{11}$, operating at higher $n$ unlocks greater connectivity, which is valuable for QEC protocols and applications\cite{Poole2024}.

For quantum computers to provide meaningful scientific and commercial value, they will likely need to support fault-tolerant operation on at least $\sim$100 logical qubits\cite{kivlichan2020improved, Campbell2021, Hoefler2023}, and neutral-atom architectures are evolving rapidly towards that goal.  This work demonstrates important new capabilities on that path, namely the ability to implement individually addressed high fidelity entangling gates and non-destructive state-selective readout, which collectively enable faster qubit operations and a broader set of QEC protocols. Ultimately, a compelling architecture for fast, fault-tolerant quantum computing is likely to integrate these techniques alongside other demonstrated neutral-atom capabilities, including dual-species platforms\cite{Anand2024, Singh2023}, mid-circuit measurement\cite{Chen2022, Lis2023, Anand2024, Singh2023, GrahamMidcircuit, Bluvstein2024}, and alternate addressing architectures\cite{Graham2022, Zhang24Thompson, PhysRevX.12.021028, Bluvstein2024}. In particular, future work should investigate the optimal combination of individual addressing, global beams, and mid-circuit rearrangement techniques to minimize circuit runtimes. For example, individual addressing can enable fast, parallel, short-range CZ gates, while motion unlocks long-range connectivity.  And while a shared global beam can support parallel, identical CZ gates, individual addressing is preferable in the case of \RZ, to support parallel gates with distinct rotation angles. Additionally, individual addressing with tighter beams provides greater flexibility in achieving the desired laser intensity at significantly lower laser power, which is an important consideration for scaling and achieving higher-fidelity gates. The advanced optical control techniques we demonstrate here also extend directly to alternative atomic species\cite{ Graham2022, Ma2023,scholl2023erasure, Bluvstein2024, rodriguez2024experimental, reichardt2024logical}, as well as multi-qubit ($N>2$) gates\cite{Levine2019}, which can further optimize logical operations\cite{Perlin2023}.

\bibliographystyle{naturemag}

\bibliography{MendeleyBib_Jan_19_2025.bib, SupplementalBib.bib}

\section{Methods}

\subsection{QPU details\label{sec:qpu}}

At the core of our platform, the qubit array is housed in a custom Infleqtion-manufactured dual-chamber ultra-high-vacuum glass cell, which lives at the center of the Quantum Processing Unit (QPU, Fig. 5). From the lower chamber, a cold atomic beam is generated with a 2D magneto-optical-trap (MOT), which loads a 3D MOT in the upper chamber. Surrounding the upper chamber are components to stabilize the electric and magnetic field environment, as well as a 405 nm light source for inducing atom desorption from the cell walls, and a microwave horn for implementing GR gates (Sec \ref{sec:implementation}). Large beams address the whole atom array for implementing the 3D MOT, polarization gradient cooling, Raman sideband cooling\cite{Tian24}, readout, and optical pumping. A 2D array of 1064~nm traps is generated by a pair of acousto-optic deflectors (AODs).  Rydberg excitation light is provided by 1040~nm and 459~nm lasers, both locked to a stable, high finesse reference cavity. The two-photon detuning from the atomic transition is actively stabilized using a ground-Rydberg Ramsey signal. For each Rydberg wavelength (1040~nm and 459~nm), separate pulse shapers (one each for control/target qubit) feed separate beam-steering modules. The pulse shapers consist of double-pass acousto-optic modulators (AOMs), optimized for fast rise times. The 1040~nm beamline additionally contains an electro-optic modulator (EOM) for implementing the phase modulation used for \CZ gates.  The beam steering modules each consist of a pair of AODs with orthogonal axes, for rastering to any location within the array. For each Rydberg wavelength, the control and target beams are combined on a non-polarizing beamsplitter.  The 1040~nm light is combined with 1064~nm light and focused onto the atoms from one side of the cell, via high numerical aperture ($\mathrm{NA}=0.7$) objective lenses, while the 459~nm light is focused from the opposing side, using a matching high-NA lens. These lenses also collect atom fluorescence during readout and direct it to an electron-multiplying CCD camera (EMCCD). Atom-based beam profiling of the 1064~nm tweezer measured a Gaussian beam waist radius at $1/e^2$ of peak intensity of 1.4(2)~$\mu\text{m}$; beam waist estimates of 2.8 and 3~$\mu$m for 459~nm and 1040~nm beams, respectively, are discussed in Sec.~\ref{sec: quantum_computer_architecture}. The trap depth is calibrated by measuring the intensity dependence of the qubit microwave transition; for the benchmarking data presented in this work, the estimated trap depth is 1~mK. The QPU is enclosed for electromagnetic, thermal, and vibrational isolation and stabilization.

\subsection{Cooling and state preparation}\label{sec:cooling_and_state_prep}
Here we give a brief description of how we cool atoms inside the optical trap and initialize them in the $|0\rangle=|F=3, m_F=0\rangle$ state. Once the atoms are loaded into the optical trap, we turn on a pair of counter-propagating beams on all three axes to provide polarization gradient cooling (PGC). The same beams are used for occupancy readout but at smaller detuning and different intensity balance. After PGC, the magnetic bias field is stepped from zero to approximately 6~G along the x-axis and a circularly polarized 895~nm laser optically pumps atoms to the stretched state $|F=4, m_F=4\rangle$. After 10 ms of optical pumping, we typically measure a temperature of $\sim4-5\,\mu\text{K}$ using the release-and-recapture method. This is followed by a Raman sideband cooling (RSC) step\cite{Tian24}, where optical pumping pulses and Raman pulses alternate. Using sideband thermometry, we measure mean trap phonon numbers of $\bar{n}_{\text{ax}} = 6.1(65)$ for an axial trap frequency of 11.5~kHz and $\bar{n}_{\text{rad}} = 0.49(22)$ for a radial trap frequency of 60.9~kHz. We believe these mean phonon numbers are currently limited by nonlinear tone modulation in the AODs producing our trap light, as changing the tone separation significantly changes the achievable temperature.  Future devices that use spatial light modulators (SLMs) should be able to obtain lower temperatures. At the end of RSC, a sequence of composite microwave pulses transfers atoms from the stretched state to qubit state $|1\rangle$. After the microwave pulse sequence is complete, the magnetic bias field is rotated from 6~G along the x-axis to the final bias field of 11.15~G along the z-axis. Once the rotation is complete, a microwave pi pulse transfer the atoms from $|1\rangle$ to $|0\rangle$, and this marks the beginning of a quantum circuit. 

The quantum circuit shot rate data discussed in Sec.~\ref{subsubsec: shot_rate_enhancement} used a different optical pumping method in which a linearly polarized 895~nm beam pumps atoms towards $|1\rangle=|F=4, m_F = 0\rangle$ at 11~G.

\begin{figure}
    \centering
    \includegraphics[width=1\linewidth]{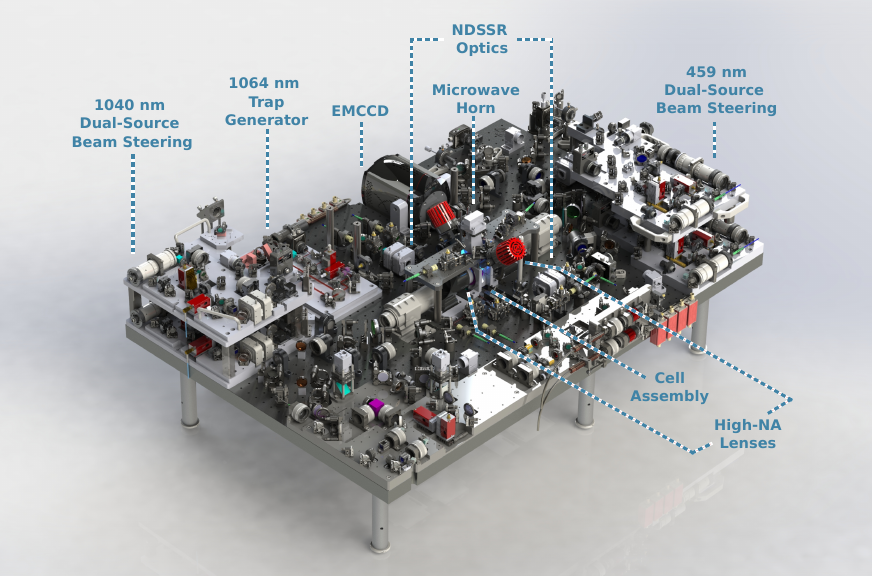}
    \caption{3D Model of the Quantum Processing Unit (QPU), highlighting the dual-source beam steering modules for both Rydberg wavelengths, a microwave horn for global \GR gates, AODs for generating a 2D array of 1064~nm optical tweezers, the electron-multiplying CCD (EMCCD) for imaging, the optics supporting non-destructive state-selective readout (NDSSR), the ultra-high-vacuum glass cell, and the high-NA lenses for focusing addressing beams and collecting readout light.}
    \label{fig:qpu_rendering}
\end{figure}

\subsection{Benchmarking of \RZ gates\label{sec:rz_characterization}}
We characterize our \RZ gates using interleaved randomized benchmarking (IRB), randomly sampling each gate over $\phi\in[0,2\pi)$.  Given the implementation of our \RZ gate, we expect a significant scattering error contribution due to off-resonant excitation to the $7P_{1/2}$ state.  This can produce leakage error that populates ground-state Zeeman sublevels outside the qubit manifold. In these states the fidelity with the target state is $F_{\rm leak}=0$, in contrast with true depolarizing error with $F_\sigma=1/2$. Denoting this leakage probability as $P_{\rm leak}$, and the depolarizing probability (conditioned on not experiencing a leakage error) as $\sigma$, we can express the probability per gate to implement the desired unitary as $C = (1 - P_{\rm leak})(1-\sigma)$.  With this definition, the data is then fit to $\frac{1}{2}+\frac{\mathrm{SPAM}}{2}C^N$, where $\mathrm{SPAM}$ is a fit parameter that accounts for reduced contrast arising from state preparation and measurement error.

We can express the fidelity $F_\mathrm{RZ}$:
\begin{equation}
    F_\mathrm{RZ} = P_{\rm leak}F_{\rm leak} + (1 - P_{\rm leak})\sigma F_\sigma + (1-P_{\rm leak})(1-\sigma)\\
\end{equation}
Using $F_{\rm leak}=0$ and $F_\sigma=1/2$, this yields
\begin{equation}
    F_\mathrm{RZ} = (1-P_{\rm leak})(1-\sigma/2)
\end{equation}

The fidelity extracted from a given cycle polarization $C$ will thus depend on the ratio of depolarization to leakage.  Since scattering error is well understood, we choose to model its contribution, and use this assumption in calculating the fidelity. We can predict the scattering error based only on the intermediate-state lifetime and detuning ($\Delta$).   For the datasets presented in Fig. \ref{fig:gr_rz_benchmarking}, we find:
\begin{center}
\begin{tabular}{ c|c|c|c|c } 
 $\Delta/2\pi$ & Fitted C & Modeled $P_{\rm leak}$ & Extracted $\sigma/2$ & $1-F_\mathrm{RZ}$\\ 
 \hline
 +885~MHz & 99.65(5)\% & 0.125\% & 0.11(3)\% & 0.24(3)\%\\ 
 -2.1~GHz & 99.87(2)\% & 0.066\% & 0.03(1)\% & 0.098(8)\%\\ 
\end{tabular}
\end{center}
We attribute the improved depolarizing error at $\Delta/2\pi=-2.1$~GHz to improved laser intensity stability.

We expect to improve the fidelity of \RZ operations with larger intermediate-state detuning. For detunings significantly greater than the ground-state hyperfine splitting, the differential light shift and photon scattering rate both scale as $1/\Delta^2$, resulting in a fundamental upper bound on the figure-of-merit — the \RZ rotation angle achieved per scattered photon — of approximately $2\pi \times 700$ radians. By comparison, at $\Delta/2\pi = -2.1$~GHz, this figure-of-merit is approximately $2\pi \times 250$ radians. It is worth noting that this upper bound can also be approached at $\Delta/2\pi = +4.6$~GHz, which is approximately half the ground-state hyperfine splitting. However, at this detuning, both the differential light shift and scattering rate are relatively large. This increased differential light shift is undesirable if the same laser is used for \RZ and \CZ operations, as it can exacerbate qubit-pair decoherence after a \CZ gate. Therefore, to minimize such coherence loss, we aim to operate at a larger red detuning to improve \RZ fidelity.

\subsection{Calibration of \CZ  gates\label{sec:cz_calibration}}
CZ gate calibration proceeds in two phases.  The first phase establishes AOM pulse parameters for resonant Rydberg excitation with the desired two-photon Rabi rate $\Omega$ and component single-photon Rabi rates $\Omega_{459}$ and $\Omega_{1040}$. Operationally, these single-photon rates are established by first setting the 459 nm laser power to give the desired differential light shift on the qubit, then varying the 1040~nm laser power to achieve the desired two-photon Rabi frequency.
This is accomplished by iterated time/frequency-domain characterizations of Rydberg Rabi excitation.  
This calibration is executed sequentially on all qubits in the array in a single circuit, allowing parallelized calibration of all qubits.

The second phase calibrates the CZ gate pulse sequence, which consists of two components: a Rydberg pulse with sinusoidally modulated phase, followed by single-qubit phase corrections required to produce the canonical \CZ gate.  
The uncorrected gate is calibrated first, by maximizing a cost function (Fig. \ref{fig:cz_benchmarking}d) that is insensitive to the single-qubit phases.  
The circuit begins and ends with $R(\theta=\pi/2, \phi=0) \equiv $\RXof[$\theta=\pi/2$] gates, surrounding N pairs of CZ gates, with an \RXof[$\pi$] gate interleaved within each pair.  
These interleaved gates provide the desired insensitivity to the single-qubit phases.
A properly tuned gate results in all population returning to $\ket{11}$ and detecting as bright-bright, while any dark signal indicates either a state $\neq\ket{11}$ or atom loss, both of which correspond to a gate imperfection.  

\begin{figure}
    \centering
    \includegraphics[width=1.0\linewidth]{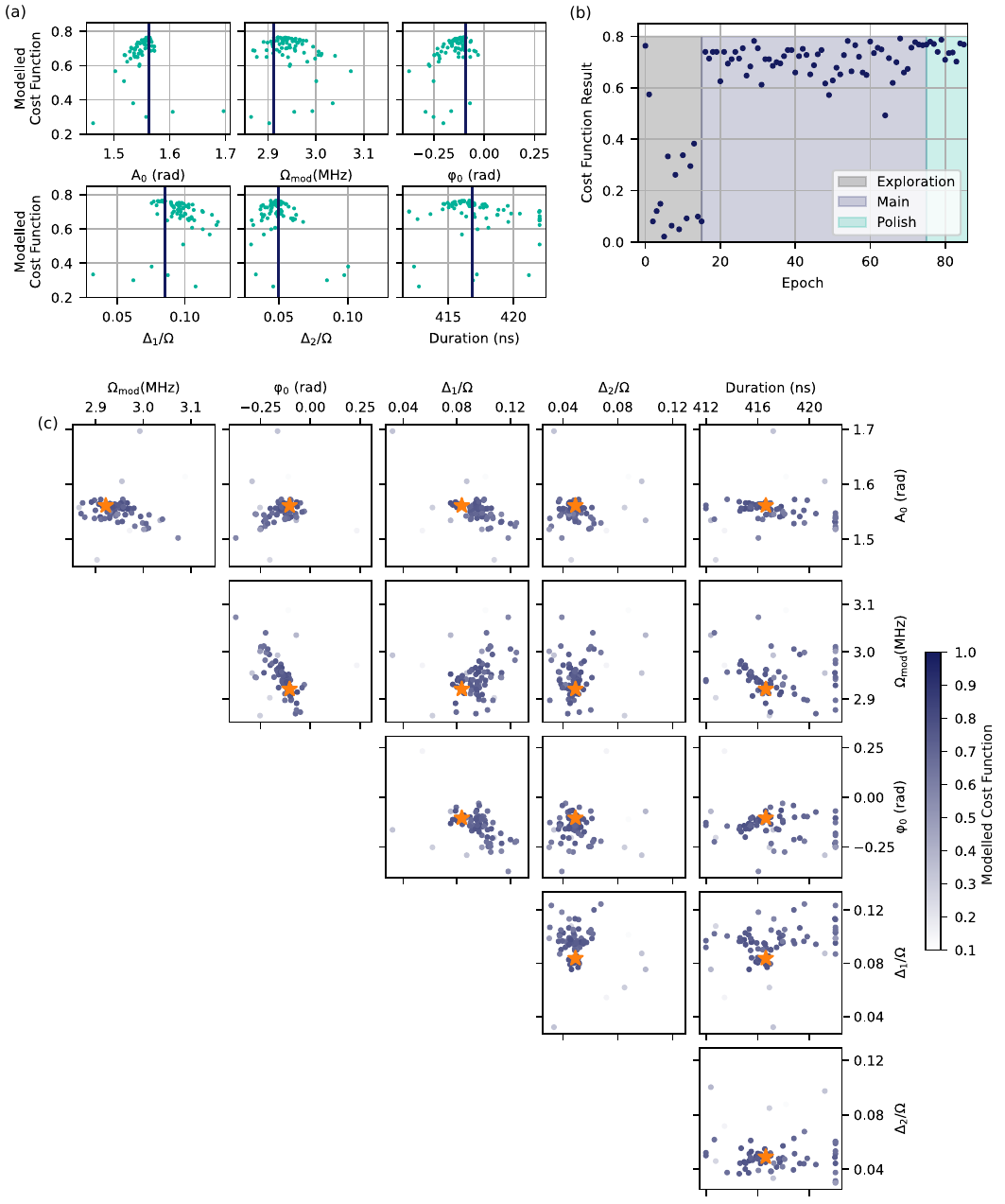}
    \caption{\textbf{Calibration of \CZ gates.} \textbf{a,} 1D projections of the final Gaussian Process model evaluated at the parameter values sampled by the optimizer. \textbf{b,} The trajectory of the measured cost function, across three different phases of the optimizer. \textbf{c,} 2D projections of the final model, evaluated at the parameter values sampled by the optimizer.}
    \label{fig:cz_BBO}
\end{figure}

Using this cost function, we optimize a representation of the gate using six parameters: the amplitude ($A_0$), frequency ($\Omega_\mathrm{mod}$), and phase offset ($\phi_0$) of the 1040~nm phase modulation, as well as the overall pulse duration and two-photon detuning.  To compensate for calibration errors, we include separate two-photon detuning corrections for each qubit, though the protocol itself should be symmetric.  To optimize these 6 parameters, we utilize a Bayesian optimizer (built on the BoTorch library) with a Gaussian process (GP) surrogate model.  This allows us to optimize without making any assumptions about the underlying parameter landscape.  Indeed, simulations suggest that our phase modulation waveform is somewhat over-parameterized, leading to the possibility of multiple local optima with similar performance. The optimizer run that preceded the benchmarking data from Fig. \ref{fig:cz_benchmarking} in the main text is shown in Methods Fig. \ref{fig:cz_BBO}.  In Methods Fig. \ref{fig:cz_BBO}b, we see the overall trajectory of the optimizer, over the course of 85 cost function evaluations, divided into an initial random sampling phase, a main exploration phase, and a final polish phase. Fig. \ref{fig:cz_BBO}a shows the final GP model, evaluated at the sampled parameter values, projected onto each parameter axis.  The same data is shown in Fig. \ref{fig:cz_BBO}c, projected onto 2D slices of the space.  This latter view can be useful for identifying correlations between parameters and finding improved parameterizations. Once an optimized waveform is identified (Fig. \ref{fig:cz_benchmarking}c), the single-qubit phases are measured via a qubit Ramsey sequence and appropriate corrections are determined.

\subsection{Benchmarking of \CZ gates\label{sec:cz_benchmarking_details}}

We compute the \CZ fidelity from the results of our benchmarking circuits by measuring the probability of the circuits to produce the expected outcomes, which in this case are uniformly $\ket{00}$. 
In order to account for the possibility for each \CZ gate to produce atom loss, leakage, or errors within the qubit subspace, we make use of the information afforded by our non-destructive state-selective readout capability (NDSSR, see Sec. \ref{sec:NDSSR}). 
By sequentially performing NDSSR and occupancy measurements, we can determine whether a given atom was retained, and if retained, whether it was in the $F=4$ (bright signal) or $F=3$ (dark signal) ground-state manifold. 
We therefore separately analyze, to a gate depth of 20, the probability to retain both atoms ($P_\mathrm{ret}$) and the probability to observe a dark-dark measurement outcome in the state-selective measurement, given that both atoms were retained ($P_{00|\mathrm{ret}}$) (see Fig. \ref{fig:cz_benchmarking}).
We fit these data to decaying exponentials with asymptotes of zero and one-quarter, respectively, and find a two-atom survival probability of $R=99.62(3)$\% per gate and a cycle polarization of $P=99.67(4)$\%. 
The observed cycle depolarization has contributions from true depolarizing error and from leakage to states $\ket{6S_{1/2}, m_F \neq0}$.  The former asymptotically produces the fully mixed state (with 25\% fidelity with the target state), and the latter produces a state that has zero overlap with the desired state. 
A detailed model for the dynamics of the intermediate-state and Rydberg populations during the gate provides an estimated leakage probability of $P_{\rm leak}=0.10$\% per gate. We use this to calculate the \CZ fidelity as $F=R \times (1-P_{\rm leak}) \times (1-\frac{3}{4} \sigma)$, with $\sigma = (1-P_{\rm leak} - P)/(1-P_{\rm leak})$ the calculated probability of depolarizing error; we find a \CZ fidelity of $99.35(4)$\%.  
Using the same leakage rate, we can calculate a \CZ fidelity post-selected for atom loss of $F= (1-P_{\rm leak}) \times (1-\frac{3}{4} \sigma)=99.73(3)\%$.

The circuits we use to benchmark \CZ gates consist of four steps: (a) preparation of the $\ket{00}$ qubit state; (b)~application of a variable number of gate pairs consisting of the same Haar-random SU(2) gate applied to both qubits, followed by a \CZ gate; (c) an inversion operation consisting of alternated application of four global \GR rotations and three \CZ gates; and (d) another layer of $N_\mathrm{pad}$ 
Haar-random global SU(2) gates followed by a final inversion operation. Steps b-d in each circuit are constructed so that, in the absence of error, the circuit performs the identity operation on the input $\ket{00}$ state. 
The final layer of random SU(2) gates is included to hold the total number of single-qubit gates constant at a value $N_\mathrm{max}$ as the depth of layer (b) is varied. The data in Fig. \ref{fig:cz_benchmarking} is a composite of two datasets, acquired concurrently, with $N_\mathrm{max}=$ 20 and 24;  we verify that this composition does not introduce significant systematic error in the extracted fidelity by simulating the full benchmarking procedure (see Sec. \ref{sec:benchmarking_simulations}).
We calculate the inversion operation in step (c) as an alternating sequence of four CZ gates and four SU(2) gates, and then we omit the final CZ gate from the circuit because it does not change measurement outcomes. The structure of benchmarking circuits is depicted in Fig. \ref{fig:cz_benchmarking}e. The global SU(2) operations are implemented using a single global \GR gate and an update to the qubit phase frame, which allows us to use only the global microwave field for the single-qubit gates in the CZ benchmarking circuits.

\subsection{Simulation of CZ benchmarking\label{sec:benchmarking_simulations}}
We conduct circuit-level simulations of our benchmarking procedure using a model for the CZ gate process. The error model includes transitions from the $\ket{1}$ state to a loss state $\ket{L}$ (no analogous term for the $\ket{0}$ state is included, because only $\ket{1}$ is near-resonantly coupled to the Rydberg state), as well as transitions between $\ket{0}$, $\ket{1}$, $\ket{0_L}$ and $\ket{1_L}$; the latter two states are leakage states that represent the $\ket{F=3, m_F\neq0}$ and $\ket{F=4, m_F\neq0}$ ground-state Zeeman sublevels and that are detected as $\ket{0}$ and $\ket{1}$ respectively. This model approximates spontaneous transition dynamics between the ground-state Zeeman sublevels, where a more complete model would separately treat all 16~ground states.

\begin{figure}
    \centering
    \includegraphics[width=\linewidth]{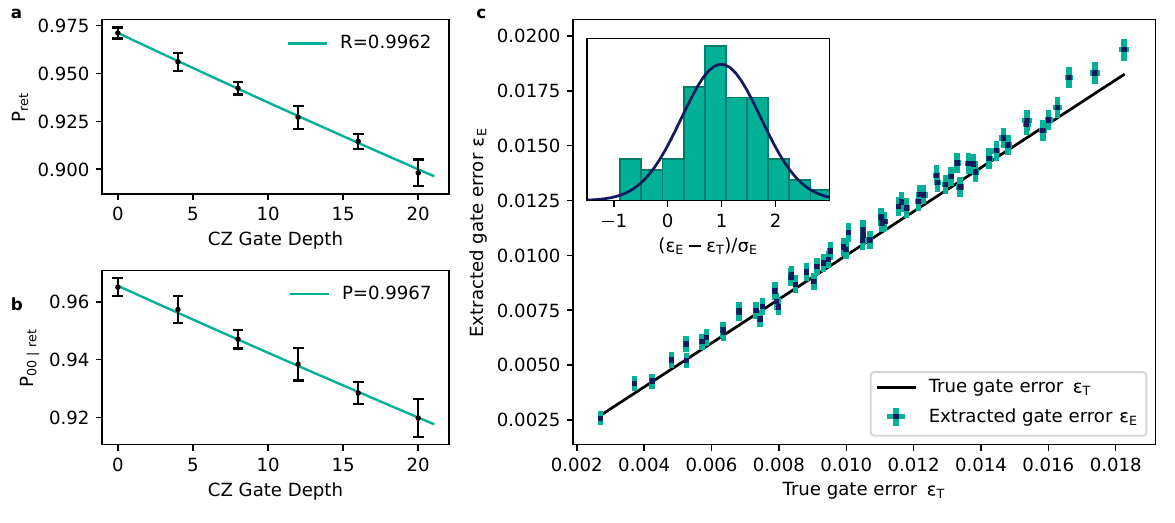}
    \caption{\textbf{Simulated \CZ benchmarking.} \textbf{a,} Predicted probability to retain both atoms in a simulated benchmarking run using the error model described in Sec. \ref{sec:benchmarking_simulations} and the circuits used to collect the data presented in Fig. \ref{fig:cz_benchmarking}; compare to Fig. \ref{fig:cz_benchmarking}f. \textbf{b,} Predicted probability to obtain the expected state-selective measurement outcome, given that both atoms were retained, in the same simulated benchmarking run; compare to Fig. \ref{fig:cz_benchmarking}g. This simulation neglects projection noise. \textbf{c,} Results of a simulation of 64 complete benchmarking runs, as described in Sec. \ref{sec:benchmarking_simulations}. This simulation includes projection noise. The figure depicts the extracted gate error and confidence interval versus the calculated gate error, obtained by averaging over Haar-random two-qubit input states. The inset histogram shows the observed distribution of the difference between these two quantities, normalized to the uncertainty in the extracted gate error in each case. The depicted Gaussian fit to this data has a mean of 0.99 and a standard deviation of 0.74, indicating that (a) our benchmarking procedure systematically underreports fidelity (by roughly 0.04\%) and (b) the extracted uncertainty somewhat overstates the statistical uncertainty.}
    \label{fig:simulated_benchmarking}
\end{figure}

We calculate spontaneous transition probabilities, accounting for (a)~interference between coupling of population from $\ket{1}$ and from the Rydberg level $\ket{R}$ to the intermediate state that produces ``dark-intermediate-state" ground-Rydberg coupling \cite{Evered2023}, (b)~interference in coupling paths through $\ket{7P_{1/2}, F=4, m_F=1}$ and $\ket{7P_{1/2}, F=3, m_F=1}$, (c)~the full decay chain from higher-lying $7P_{1/2}$ decay products $5D_{3/2}$ and $7S_{1/2}$, and (d)~Rydberg decay-to-ground events. The result is a matrix describing the probability of a spontaneous transition per gate, expressed in the $\{\ket{0}, \ket{1}, \ket{0_L}, \ket{1_L}\}$ ordered basis:
\begin{align}
    \begin{pmatrix}
        17.4 & 185.0 & 4.9 & 165.4 \\ 18.5 & 197.5 & 4.6 & 177.7 \\ 31.1 & 420.3 & 45.8 & 1210.0 \\ 42.1 & 590.2 & 52.9 & 1901.0
    \end{pmatrix} \times 10^{-6}.\label{eq:leakage_matrix}
\end{align}
Our error model includes these transition probabilities, as well as a probability per gate to transition from $\ket{1}$ to $\ket{L}$ of 0.39\% and an additional single-qubit $Z$ error probability of 0.065\%. These latter two probabilities are chosen such that a simulation of the circuits used to obtain the benchmarking data shown in Fig. \ref{fig:cz_benchmarking} exactly reproduces the observed retention and polarization probabilities per gate. The results of this simulation are shown in Fig. \ref{fig:simulated_benchmarking}. The choice of $Z$ as the only Pauli error is well-motivated by the underlying physics, as the dominant source of bit-flip errors is already embodied in the scattering matrix of Eq. \ref{eq:leakage_matrix}. 

By employing both first-principles and phenomenological inputs, this model provides valuable validation of the benchmarking routine and its analysis. For example, the true asymptote of the probability to observe the dark-dark outcome in the state-selective measurement is certainly different from the  25\% that is used for the fit to the data. This difference arises from the depumping from the $F=4$ hyperfine level driven by the 459 nm laser. Nevertheless, we calculate gate fidelities averaged over Haar-random input states and over twelve symmetric stabilizer states that form a 2-design on the two-qubit symmetric subspace \cite{Tsai2024} to be 99.38\% and 99.37\%, respectively, for the error model described above that reproduces the experimental observations. These fidelities are greater than and consistent with the 99.35(4)\% fidelity extracted from the benchmarking data for both the experimental and simulated datasets. These comparisons suggest that our benchmarking routine adequately characterizes the gate for our gate physics.

To further validate our benchmarking approach, we compare the gate fidelity averaged over Haar-random input states to the fidelity extracted from simulated benchmarking for a variety of error models. These models are obtained from the one thus far described by independently scaling the leakage matrix in Eq. \ref{eq:leakage_matrix}, the $\ket{1}\rightarrow\ket{L}$ probability, and the single-qubit dephasing probability. Benchmarking runs are simulated with the same number of circuits at each depth, and with the same $N_\mathrm{max}$, as used to obtain the data shown in Fig. \ref{fig:cz_benchmarking}, and the simulated analysis routine includes projection noise. Across these runs, the extracted error $\varepsilon_\mathrm{E}$ is larger than the true error $\varepsilon_\mathrm{T}$ in 86\% of runs, with an average difference $\varepsilon_\mathrm{E}-\varepsilon_\mathrm{T}=0.04$\%. We present further quantitative details in Fig. \ref{fig:simulated_benchmarking}c.

\subsection{Impact and mitigation of optical crosstalk in the scanned-beam addressing architecture\label{sec:crosstalk_simulations}}

In our addressing geometry, the lasers that drive Rydberg excitation propagate perpendicular to and are focused in the plane of the atomic qubit array, across which their foci are scanned using acousto-optic deflectors. System level performance considerations (e.g. increasing qubit count and connectivity) exert pressure to reduce spacing in atom arrays. While in principle addressing beams' spatial profiles can be shaped to effectively eliminate intensity on non-target atoms (which we refer to as ``crosstalk"), in practice crosstalk elimination is challenging over large optical field of view, for optical systems consisting of many optical surfaces, and for atom separations that approach the optical wavelength. For example, given 6 $\mu$m qubit spacing and 3 $\mu$m spot radii for Gaussian Rydberg beams, one predicts an intensity crosstalk $\eta$, the ratio of intensity on a neighboring spectator site to peak intensity, of $0.03$\%. In practice atomic measurements suggest $\eta$ up to 10 times larger in the worst case, likely due to the impact of wavefront distortions introduced by optical surface imperfections. The practical inevitability of crosstalk motivates a discussion of two of its primary effects and their relevance to high-fidelity and fault-tolerant operation of our neutral-atom quantum processor.

\subsubsection{Effect of optical crosstalk on entangling gate error}
In an entangling gate multiple atoms are addressed by different fields, and crosstalk error contributions fundamentally arise due to the interference between the direct addressing field and the crosstalk field on each atom. If this interference can be compensated, for example by appropriately modulating the intensities of the fields, then the error contribution vanishes. However, such compensation presents a challenging control problem, suggesting that a different solution is more promising: we take an approach where the frequency of this interference $\omega_X$ can be made much larger than the Rabi frequency $\Omega$ describing the gate dynamics. Simulations confirm that as the ratio $\omega_X/\Omega$ increases, the Rabi dynamics average over this increasingly fast perturbation, allowing the desired gate dynamics to be realized. They also reveal that achieving a ratio of $\omega_X/\Omega$ of about 50 is sufficient, for the particular crosstalk intensities we investigate, to make the CZ error contribution from optical crosstalk negligible compared to error correction thresholds.  Our system currently operates at $\omega_X/\Omega\approx5$, but with modifications to the frequency/diffraction-order of the pulse shaping AOMs, it is straightforward to reach $\omega_X/\Omega>50$.
This modeling suggests that maintaining a sufficiently large beatnote between neighboring addressing beams is a straightforward and effective way to mitigate crosstalk in an individual-addressing architecture.  Future work could explore alternative mitigation strategies such as operation at zero beatnote frequency with active or passive stabilization of the relative phase between the two beams.

\begin{figure}
    \centering
    \includegraphics[width=1.0
\linewidth]{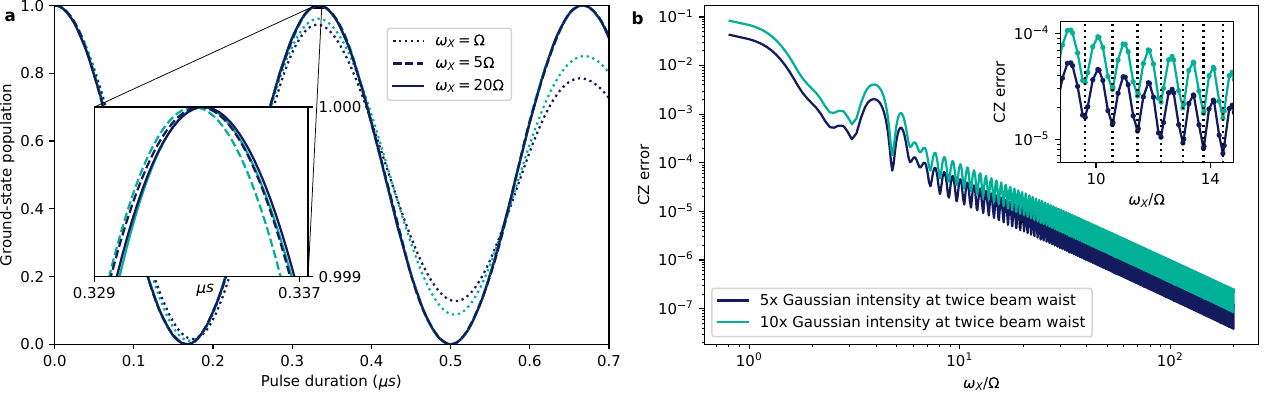}
    \caption{\textbf{Optical crosstalk effect on CZ gates. a,} Representative simulated trajectories for Rabi oscillations between the $6S_{1/2}$ ground state and the $69S_{1/2}$ Rydberg level, including the effect of optical crosstalk from Rydberg-excitation light addressing a neighboring site. The crosstalk intensity used in the simulation is 5 $\times$ higher than the intensity 6 $\mu\rm m$ from the center of a Gaussian beam with 3 $\mu\rm m$ waist radius at $1/e^2$ of peak intensity (see main text). An important parameter is the crosstalk frequency $\omega_X$ relative to the frequency $\Omega=2\pi\times3$ MHz of Rabi oscillations, and three values for this parameter are shown: $\omega_X/\Omega=1$ (dotted trajectories), 5 (dashed trajectories), and 20 (solid trajectories). For each crosstalk frequency we display two trajectories (distinguishable by color), with different phase relationships between the crosstalk fields for each trajectory, highlighting the sensitivity to this phase relationship for $\omega_X/\Omega=1$. Simulations include both Rabi frequency and light-shift oscillations that arise due to optical crosstalk, and include no other error sources. \textbf{b,} Calculated CZ error as a function of the ratio $\omega_X/\Omega$. The error can be made smaller by increasing this ratio. At ratios greater than about 6 regular structure appears, with error minima when the gate duration is an integer multiple of the crosstalk period. The inset shows this structure over a smaller range, with vertical lines indicating the frequencies at which this integral relationship is satisfied; it also emphasizes our discrete sampling of the ratio $\omega_X/\Omega$ in the simulation. We find that the gate error is approximately proportional to the crosstalk intensity, for which we depict two values here.}
    \label{fig:crosstalk_simulations}
\end{figure}

We depict simulations of the impact of optical crosstalk on single-atom Rabi oscillations to the Rydberg state and on CZ error in Fig. \ref{fig:crosstalk_simulations}.

\subsubsection{Effect of optical crosstalk on spectator qubits}
During implementation of an \RZ or \CZ gate, the addressing lasers irradiate spectator qubits with some small crosstalk intensity. In the case of an \RZ gate, crosstalk from the 459 nm laser produces an unintended \RZof[$\phi_X$] gate on spectator qubits, where $\phi_X=\eta\phi$ for relative crosstalk intensity $\eta$. The realized \RZof[$\phi_X$] rotation implements the desired identity operation on the spectator qubit with an expected error $\epsilon_X=\frac{2}{3}\sin^2{\left(\phi_X/2\right)}$ for a Haar-random input spectator state. This error is quadratic in the crosstalk intensity $\eta$. As an example, if we take $\eta=0.003$ as an approximate description of our current system, for $\phi=\pi$ we have $\epsilon_X = 1.5\times10^{-5}$.

In the case of a CZ gate, the 459 nm laser applies a similar undesired \RZof[$\phi_X$] rotation to spectator qubits. For our current parameters, the rotation applied to a spectator qubit by a CZ gate is equivalent to that from an \RZof[$\phi=1.7\pi$] gate on the target qubit, for which we calculate $\epsilon_X=4.3\times10^{-5}$ for $\eta=0.003$. 

The impact of these spectator-qubit rotations on overall circuit fidelity is circuit dependent. We estimate the impact using simple assumptions: that every qubit is targeted by the same number of \RZ gates, and that every nearest-neighbor pair is targeted by the same number of CZ gates. We calculate an effective adjustment to the fidelities of the \RZ and CZ gates on a given atom by including error accumulated during other \RZ and CZ gates in which that atom is a nearest-neighbor spectator. In a square lattice under this assumption, the ratio of nearest-neighbor-spectator/target \RZ rotations a given qubit experiences is 4, while for CZ gates this ratio is 3. We can therefore add an effective, heuristic error contribution to the \RZ gate error budget of $4\times 1.5\times 10^{-5}=6\times10^{-5}$. To the CZ gate error budget we add an analogous $2\times3\times4.3\times10^{-5}=2.6\times 10^{-4}$, where the first factor of 2 arises from the number of qubits participating in the gate. This suggests that reducing the crosstalk intensity may be important for future fault-tolerant operation, but a more complete analysis should calculate the circuit-specific impact. Another possibility that should be explored is crosstalk-aware compilation \cite{Campbell2023}, in which the native gate set contains gates not of the form $R_Z(\phi)\otimes I\otimes...\otimes I$, but instead $R_Z(\phi)\otimes R_Z(\eta_1\phi) \otimes ... \otimes R_Z(\eta_n\phi)$ and similar for the CZ gate.

There are two more effects arising from the CZ gate that we consider: compared to the 459 nm beam, the differential qubit shift from the 1040~nm Rydberg-excitation beam is smaller by a factor larger than $10^3$, so we neglect the very small rotation applied by the 1040~nm crosstalk. Finally, coupling to the Rydberg level may arise from illumination by 1040~nm crosstalk and 459 nm crosstalk. If we assume the same relative crosstalk intensity $\eta$ for both colors, the Rabi frequency of this coupling is $\eta\Omega$, where $\Omega$ is the Rabi frequency for the target qubit. Additionally, the spectator atom's ground-to-Rydberg transition frequency differs from the target atom's by (approximately) the value of the light shifts applied to that transition by the Rydberg excitation beams. This produces a ground-to-Rydberg coupling on the spectator atom that is (a) weaker by a factor of $\eta$ and (b) very far detuned. Concretely, for our current system parameters and $\eta=0.003$, we estimate that this coupling is relatively detuned by an amount $\Delta_X/\Omega_X=10^3$. The time-averaged spectator-qubit population in the Rydberg state arising from this coupling is $5\times10^{-7}$; we therefore fully neglect this effect as well.

\subsection{NDSSR implementation}\label{sec: ndssr_methods}
\subsubsection{NDSSR sequence and optimization}
Our implementation of non-destructive state-selective readout improves upon the setup demonstrated in~\cite{Kwon2017}. At the end of circuit execution, the bias field is rotated from the z-axis to the x-axis (see Fig.~\ref{fig:qpu_diagram}c for orientation) along which two counter-propagating, circularly-polarized, near-resonant 852~nm beams optically pump atoms in the upper hyperfine manifold $F = 4$ to the stretched state $|F=4, m_F=4\rangle$ and scatter photons primarily along the closed transition $|F=4, m_F=4\rangle \rightarrow |F=5', m_F=5'\rangle$. The beams are red-detuned from zero-field resonance by $2\pi \times 1.6$~MHz, with one beam further detuned by $2\pi \times 3$~kHz to wash out the standing-wave intensity pattern; the peak beam intensity per beam is estimated to be ${\sim}1$~$\text{mW}/\text{cm}^2$, which is close to the saturation intensity of the transition (1.1~$\text{mW}/\text{cm}^2$). Atoms that are in the lower hyperfine manifold $F =3$ stay dark during this process. The intensities of the near-resonant beam-pair and the trap laser are modulated out-of-phase, so that the trap laser intensity is minimized while the atoms have a high probability of being excited by the imaging beams. The trap intensity modulation has a nominal duty cycle of 75\% and the imaging beam pair has a nominal duty cycle of 15\%. The duty cycle and the phase offset of the intensity modulation waveforms are controlled by the DDS cards driving the acousto-optic modulators for the lasers. Because there is no three-dimensional cooling, with only Doppler cooling expected on the imaging beam axis, atom temperature quickly rises during this readout and hence exposure time has to be limited such that atom loss from the trap is minimized while achieving good discrimination between bright and dark distributions of the camera signal. Exposure time of 6~ms was found to be sufficient to provide high bright-dark discrimination fidelity (99.6(2)\%). By carefully optimizing the laser power modulation parameters, power balance between counter-propagating beams, and the bias field, we were able to achieve state-average atom-loss probability of ${<}1$\% ( Fig.~\ref{fig:ndssr_main_figure}b) in a time-averaged trap depth of 1.97 mK (peak trap depth 2.6 mK), which is significantly lower than  ${>}10$~mK used in \cite{Kwon2017} and \cite{Nikolov2023} while achieving better atom loss rate. Preliminary results of further optimization and technical improvements not discussed in this work suggest that the peak trap depth can be lowered to 2~mK while maintaining the state discrimination fidelity and atom loss rate mentioned in Sec.~\ref{sec:NDSSR}. In particular, we observed an interesting correlation between the optimal intensity modulation frequency for minimal atom loss and the Zeeman energy difference between $m_F$ levels of the upper-hyperfine manifold, which warrants further investigation; for the readout sequence used in this work, we chose a x-bias field of 3.59~G and intensity modulation frequency of 1.32~MHz. Leakage of off-resonant repump laser light due to the finite extinction ratio of acousto-optic switches can result in increased dark-to-bright state leakage at deep trap depths, which accounts for some of the 1.6(5)\% dark state discrimination error mentioned in Sec.~\ref{sec:NDSSR}.

\subsubsection{Discrimination fidelity modeling}
\begin{figure}[t]
\centering
    \includegraphics[width=89mm]{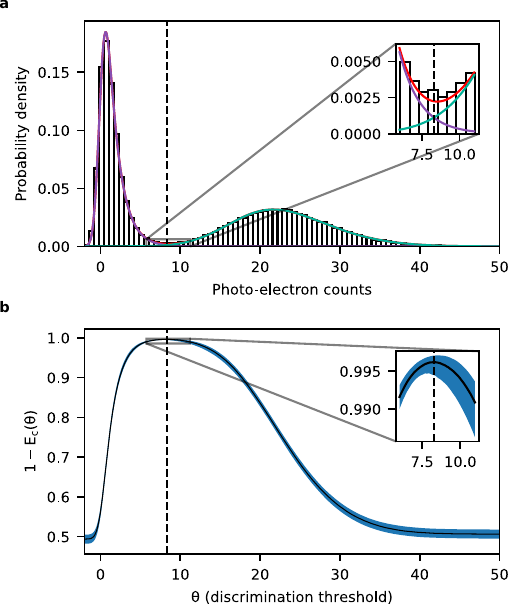}
    \caption{\textbf{NDSSR discrimination fidelity modeling.} The black-dashed line denotes the optimal discrimination threshold of 8.29 photo-electron counts. \textbf{a}, Photo-electron count distribution from the dataset used in Figure~\ref{fig:ndssr_main_figure}. The red line is a fit to a model that is a mixture of an exponentially modified Gaussian distribution (purple) and a skew-normal distribution (turquoise). The zoomed inset shows the sum of the two component distributions (red) around the optimal discrimination threshold, where there is a small overlap. \textbf{b}, Discrimination fidelity estimate as a function of the chosen threshold $\theta$ (dark line) with 1$\sigma$ confidence interval (blue band) obtained from bootstrapping. The zoomed inset shows the discrimination fidelity estimate around the optimal discrimination threshold.} \label{fig:ndssr_dark_bright_discrimination}
\end{figure}

\begin{figure}[t]
    \centering
    \includegraphics[width=0.75\linewidth]{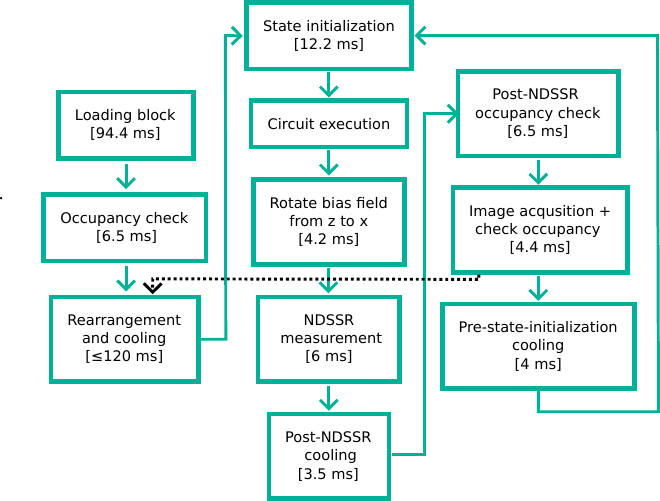}
    \caption{\textbf{NDSSR control flow.} Control flow of an NDSSR experiment cycle, with approximate duration for each step. At the step with green solid arrow and dotted black arrow, if an atom loss is detected from the qubit array or if the atom reservoir does not have enough resources for rearrangement, the control sequence follows the dotted black arrow. Otherwise, the green arrow is followed. When the required number of measurements has been collected or the atom reservoir runs out of atoms, the control sequence segues into the monitor block (not shown, duration $\sim 10\,\text{ms}$) to check laser intensity and lock status. Once the monitoring data is gathered and processed over the network ($\sim 20\,\text{ms}$), feedback corrections are applied to actuators and another cycle is launched. If the Raman sideband cooling step is enabled, an extra step is inserted between the state initialization and the circuit execution step, which effectively adds an extra 52~ms duration with our existing parameters.}
    \label{fig:ndssr_control_flow}
\end{figure}

We define discrimination fidelity as the probability of correctly classifying the photo-electron counts as coming from atomic fluorescence. For each qubit trapping site, we define a mask for selecting relevant camera pixels and then compute a weighted sum of raw camera analog-digital units (ADU) for those pixels (note each pixel has a fixed offset, which must be subtracted); the sum is then converted into photo-electron counts by dividing the sum by the EM gain (300) and then multiplying by the pre-amplifier gain (5.16 electrons per ADU count) provided by the EMCCD vendor datasheet. We model the distribution of photo-electron counts $P(x)$ as a mixture of two distributions: exponentially modified Gaussian for the dark distribution $P_D(x)$ weighted by the dark counts fraction $A_D$, and skew-normal for the bright distribution $P_B(x)$ weighted by the bright counts fraction $1 - A_D$:
\begin{align}
    P(x) &= A_{D}P_{D}(x) + (1-A_{D})P_{B}(x)\\
    P_{D}(x) &= \frac{1}{2K\sigma_D} \exp\left(\frac{1}{2K^2} - \left(\frac{x-\mu_D}{K\sigma_D}\right)\right)\mathrm{erfc}\left(-\frac{(x-\mu_D)/\sigma_D-1/K}{\sqrt{2}}\right)\\
    P_{B}(x) &= \frac{1}{\sqrt{2\pi}\sigma_B}\exp\left(-\frac{(x-\mu_B)^2}{2\sigma_B^2}\right)\left(1 + \mathrm{erf}\left(a\frac{x-\mu_B}{\sqrt{2}\sigma_B}\right)\right)
\end{align}
The dark distribution is parametrized by $\mu_D$, $\sigma_D$, and $K$, which describe the location, scale, and shape of the dark distribution, respectively. The bright distribution is parametrized by $\mu_B$, $\sigma_B$, and $a$, which describe the location, scale, and shape of the bright distribution, respectively; $\mathrm{erf}(x)$ and $\mathrm{erfc}(x)$ are the error function and the complementary error function. We obtained the following parameter values after fitting the proposed mixture distribution to the SPAM image dataset used in Fig.~\ref{fig:ndssr_main_figure}: $A_D = 0.508, \mu_D = -0.007, \sigma_D = 0.600, K=2.363, \mu_B = 16.897, \sigma_B = 8.894, a = 1.802$ (see Fig.~\ref{fig:ndssr_dark_bright_discrimination}a for the fit plot). Regarding the choice of using exponentially modified Gaussian distribution for the dark distribution, we believe that a convolution of normal and Gamma distributions is more appropriate, based on the observation that the EMCCD ADU count per pixel distribution in the dark can be modeled as a mixture of normal (from readout noise) and Erlang distributions (from parallel clock-induced charge noise)~\cite{Bergschneider2018}, but we found that exponentially modified Gaussian distribution fits the data just as well and is easier to compute; skew-normal distribution is chosen for the the bright distribution for a similar reason of convenience.

Once a mixture distribution is fitted to the data, the optimal discrimination threshold $\theta^*$ is then determined as the photo-electron count minimizes the sum of classification errors $E_{C}(\theta)$, which is
\begin{align}
E_{C}(\theta) = A_{D}\int_{\theta}^{\infty} P_{D}(x)\,dx + (1-A_D)\int_{-\infty}^{\theta} P_{B}(x)\,dx
\end{align}
To estimate the discrimination fidelity $1-E_{C}(\theta^*)$ using this model, we applied bootstrapping to the SPAM image dataset used in Fig.~\ref{fig:ndssr_main_figure}: we created 1000 SPAM image subsets, with each subset containing 2000 images randomly sampled from the main dataset with replacement, fit the mixture distribution for the photo-electron counts in each subset and then computed $E_{C}(\theta)$. The bootstrapping result (Fig.~\ref{fig:ndssr_dark_bright_discrimination}b) shows that the discrimination fidelity saturates at 99.6\% around the discrimination threshold of 8.29 photo-electron counts.

\subsubsection{Shot rate enhancement from repetitive measurements}\label{subsubsec: shot_rate_enhancement}
In our architecture, a measurement cycle is defined by a sequence of repetitive measurements that continues until the maximum measurements allowed per cycle is reached or the atom reservoir is depleted. The shot rate enhancement effect of repetitive measurement was measured by averaging the total execution times of a quantum circuit that prepares a 15-qubit GHZ state with CZ gates for entangling, local \RZ gate for qubit phase correction and global \GR gates for applying Hadamard and dynamical decoupling gates. The test circuit consisted of 14 \CZ gates, 14 \RZ gates and 29 \GR gates. The circuit was repeated three times with each instance requiring at least 400 valid measurements and maximum measurements per measurement cycle limited to seven, and recorded total execution times of 47.7~s, 50.3~s, and 48.1~s which results in an average shot rate of 8.2(2) Hz. In this context, a raw measurement is invalidated if the post-NDSSR occupancy measurement detected an atom loss in the qubit array, and multiple measurements in a measurement cycle are invalidated if analog (laser intensity) and digital (laser lock status) validators detected samples out of validation bounds at the end of a cycle; the total execution time is defined by the time interval between the beginning of pulse sequence compilation and control device initialization and the end of the job data (camera image set) upload to the database. It should be noted that this data was taken when the \CZ gate fidelity was significantly lower (average SPAM-corrected Bell state preparation fidelity was 93.1\%) than the best \CZ gate fidelity demonstrated in this work and the NDSSR bright state atom loss rate was slightly higher ($\sim 1.5$\% instead of 1\%), resulting in higher atom loss and incurring more measurement invalidation and atom loading and arrangement iterations. We did not repeat the circuit execution for maximum one measurements per measurement cycle, but null circuit (empty circuit) measurements on the same day suggests the shot rate would have been limited to 3.2~Hz even in the absence of gate-induced atom loss.

\subsection{Acknowledgements}
We gratefully acknowledge contributions from Alan Doak, Alex Olivas, Brenden Villars, Bryna Guriel, Cody Mart, Chris Wood, David Owusu-Antwi, Denny Dahl, Emma Brann, Evan Salim, Jan Preusser, Jean-Philippe Feve, John Sivak, Jonathan Cohen, Jorge Gomez, Josh Behne, Josh Cherek, Karen Rosenthall, Marie Grubb, Marissa McMaster, Matt Ebert, Rob Williamson, Robin Keus, Ryan Prior, Salahedeen Issa, Stephanie Lee, Steven Kitchen, and Steven Sedig.

\subsection{Author contributions }

AGR, DM, TWN, WCC, DMG, AMS, DCC, MCG, MTL, GTH, ILB, TO, EC, MS, MY, SDS, AJF, RAJ, KWK, FM, IVV, MLW, MI, PTM, JRT, MKD, JRG, MRB, JKM, AR, and TMG contributed to the design and construction of the apparatus hardware,
AGR, DM, TWN, WCC, AMS, NGP, DCC, TGB, MTL, GTH, DG, ILB, TBL, CC, PG, VO, PG, EC, MJB, MEB, AKT, EBJ, RAJ, KWK, SYE, DAB, CPS, IVV, JJM, PTM, JRT, MKD, JRG, NANM, CJGJ, PJR, LLH, PDB, RR, and MAP contributed to the control system and software stack. AGR, DM, TWN, WCC, DCC, MTL, GTH, CC, PG, VO, EC, EBJ, RAJ, KWK, IVV, PTM, JRT, MKD, and JRG contributed to
operating the apparatus and collecting the data. AGR, DM, TWN, WCC, DMG, DCC, MTL, GTH, ILB, CC, PG, VO, TO, PG, JMG, EC, MS, EBJ, JDM, RAJ, KWK, IVV, JAM, PTM, JRT, MKD, JRG, NANM, AC, TMG contributed to analysing data, AGR, DM, TWN, WCC, DCC, MTL, GTH, CC, EC, MS, MY, EBJ, JDM, KWK, IVV, JRT, JRG, RR, TMG, and MAP contributed to theoretical analysis and simulation.
The manuscript was written by AGR, DM, TWN, WCC, DCC, and MS. The project was supervised by AGR, TWN, WCC, DCC, DG, PG, TO, MEB, MS, JJM, and MI. 

\subsection{Data availability}
The data presented here are available from the corresponding author on reasonable request.

\subsection{Competing interests}
AGR, DM, TWN, WCC, DMG, AMS, NGP, DCC, MCG, TGB, MTL, GTH, DG, ILB, TBL, CC, PG, VO, TO, PG, JMG, EC, MJB, MEB, MS, AKT, MY, SDS, EBJ, AJF, JDM, RAJ, KWK, SYE, DAB, FM, CPS, IVV, JJM, MLW, JAM, MI, PTM, JRT, MKD, JRG, NANM, MRB, CJGJ, PJR, JKM, AR, LLH, PDB, RR, AC, MAP are employees and/or shareholders of ColdQuanta, Inc d.b.a. Infleqtion.

\end{document}